`

# Native defect assisted enhanced response to CH$_4$ near room temperature by Al$_{0.07}$Ga$_{0.93}$N nanowires

Santanu Parida,[a,*] A. Das,[a] Arun K. Prasad,[a] Jay Ghatak,[b] and Sandip Dhara[a,*]

[a] Surface and Nanoscience Division, Indira Gandhi Centre for Atomic Research, Homi Bhabha National Institute, Kalpakkam-603102, India

[b] Chemistry and Physics of Materials Unit, International Centre for Materials Science, Jawaharlal Nehru Centre for Advanced Scientific Research, Jakkur, Bangalore-560064, India

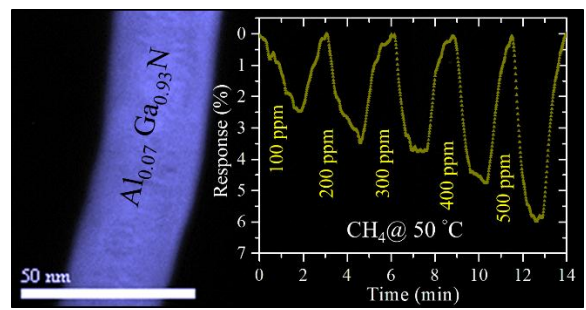

**Abstract**: Gas sensors at low operating temperature with high sensitivity are the demand for the group III nitrides owing to their high chemical and thermal stability. The CH$_4$ sensing is realized for the Al$_{0.07}$Ga$_{0.93}$N nanowires (NWs) with an improved response over the GaN NWs at a low operating temperature of 50 °C, for the first time. Al$_{0.07}$Ga$_{0.93}$N NWs were synthesized via ion beam mixing process using inert gas ion irradiation on the bilayer of Al/GaN NWs. The sensing mechanism is explained with the help of native defects present in the system. The number of shallow acceptors created by Ga vacancy (V$_{Ga}$) is found to be higher in Al$_{0.07}$Ga$_{0.93}$N NWs than those in the as-grown GaN NWs. The role of O antisite defect (O$_N$) for the formation of shallow V$_{Ga}$ is inferred from photoluminescence spectroscopic analysis. These native defects strongly influence the gas sensing behaviour resulting in the enhanced and low temperature CH$_4$ sensing.

[*]Corresponding author email: santanuparida026@gmail.com; dhara@igcar.gov.in

# 1. Introduction

Group III nitrides, the direct bandgap semiconductors, have attracted the scientific community as well as commercial sector because of their widespread application in optoelectronics. These binary semiconductors (InN, GaN, and AlN) along with their tertiary and quaternary alloys can span the wide range of electromagnetic radiation starting from infrared to deep UV by proper alloy formation.[1,2] Owing to this characteristic, III-V nitrides are used in varieties of applications like photodetector, light emitting diode, solar cell, and sensors.[3-6] In III-V nitrides, native defects play a crucial role in determining the carrier concentration and thereby modifying the electrical and optical properties of the solid.[7,8] Along with controlling the optoelectronic properties, these defects also influence gas sensing by controlling the conduction process, especially for nanostructures.[9] Moreover, the III-V nitrides are found to be chemically and thermally stable and biocompatible.[7] Therefore, it is a better choice for the gas sensing device in regard to their stability unlike that of the metal oxides.[10,11] It is well known that metal oxides show variation in stoichiometry with time in ambient conditions and hence lose the stability resulting in the fluctuation of response. The sensitivity of the device, which is reported to be very high for metal oxide nanostructure,[10-14] can be compromised at the cost of the stability of the nitride-based semiconductor for long term applications.

There are only a few reports of gas sensors based on the III nitride semiconductors. The gas sensing behaviour of the III-V nitrides is demonstrated mostly through functionalization or decorating by metal nanoparticles like Pd and Pt, where the sensitivity can be tuned by controlling the catalyst particle size and distribution.[6,15-17] The possibility of degradation of the device performance due to the poor thermal stability of Pd like metal nanoparticles is reported.[18] Moreover, gas sensing behaviour is also realized with the formation of hetero-structures using the nitride semiconductors. Generally, two types of heterostructures are proven to be a good candidate for gas sensor application, (i) the metal-semiconductor junction (Schottky diode),[19,20] and (ii) semiconductor-semiconductor junction for the formation of high electron mobility transistor (HEMT) structure,[21] especially in the III-V nitrides. In the case of Schottky diode, the sensing behaviour is explained with the help of variation in the barrier height by exposing the analyte gas. On the other hand, the formation of two-dimensional (2D) electron gas at the interface of $Al_xGa_{1-x}N$ based heterostructures provides high mobility leading to the fast response and recovery of the device in the HEMT like structure. Nevertheless, the making of heterostructures and the sharp interface between the two semiconductors are challenging and need sophisticated epitaxial growth techniques.

The above-discussed techniques and structures are mostly utilized for the sensing of $H_2$ gas. Leaving metal oxide based sensors,[10,11,22] there are very few reports on the sensing of $CH_4$ using the nitride-based semiconductor.[9] $CH_4$ is the second highest contributor to global warming with a global warming potential of 23 as compared to $CO_2$ of value 1. The detection of $CH_4$ at temperatures lower than 350 °C is difficult owing to its thermodynamic stability as compared to other hydrocarbons.[23]

Thus, much attention is paid to the research community for the sensing of $CH_4$ gas at lower temperatures, particularly near room temperature. The low-temperature sensing of $CH_4$ is realized in the metal oxide based semiconductor nanostructure because sufficient amount of activation energy can be provided by the nanostructures for triggering the chemisorption in the gas sensing process.[11,22] Whereas, there is hardly any report for the low-temperature $CH_4$ sensing behaviour using III-V nitrides. Further, obtaining high sensitivity and low operating temperature for the nitride-based gas sensors are challenging and need to be explored further.

In the present study, we demonstrate the $CH_4$ gas sensing behavior of the $Al_xGa_{1-x}N$ nanowires (NWs), synthesized in the ion beam mixing process using CVD grown GaN. Detailed studies of $CH_4$ sensing by GaN NWs at elevated temperature is described elsewhere,[9] and the primary focus is therefore on the investigation of the $CH_4$ response of $Al_xGa_{1-x}N$ NWs with emphasis on the underlying mechanism for the same. A comparison of $CH_4$ gas sensing behavior of the $Al_xGa_{1-x}N$ NWs with that for CVD grown pristine GaN NWs is presented. The higher response of $Al_xGa_{1-x}N$ NWs towards $CH_4$ than that for GaN NWs is explained considering the presence of native defects in the former, as evident from the extensive study of defects using photoluminescence (PL) spectroscopy in these nanostructures.

## 2. Experimental details

### 2.1 Synthesis of $Al_xGa_{1-x}N$ nanowires

GaN NWs were grown by the atmospheric pressure chemical vapor deposition (CVD) technique in the catalyst-assisted vapor-liquid-solid (VLS) process. The detail of the growth process is described elsewhere.[24] In brief, we used Ga metal (99.999%, Alfa Aesar) as a precursor, $NH_3$ (5N pure) as a reactant, and $N_2$ (5N pure) as the carrier gas for the growth of NWs on Si (100) wafer with Au nanoparticles as a substrate. The growth was carried out at 900 °C for a duration of 3 h. GaN NWs, grown by CVD technique, were used for the synthesis of $Al_xGa_{1-x}N$ NWs in the process of ion beam mixing (IBM). In the IBM process, as-grown GaN coated with 25 nm of Al (*via* thermal evaporation) was irradiated by $Ar^+$ at a fluence of $5\times10^{16}$ ions/cm$^2$. Post-irradiation annealing was carried out at a temperature of 1000 °C in $N_2$ (5N pure) atmosphere for 5 min as a final step for the formation of a random alloy, $Al_xGa_{1-x}N$. The detailed experiments were reported in our earlier study.[25] The fluence and post-irradiation parameters were chosen from optimum conditions obtained in the report.

### 2.2 Characterization

The vibrational properties were studied using Raman spectroscopy (inVia, Renishaw, UK) with $Ar^+$ laser excitation of 514.5 nm in backscattering geometry. A grating with 1800 grooves·mm$^{-1}$ was used as a monochromator. Resonance Raman and PL spectra were recorded at room temperature using a UV laser of wavelength 325 nm in the same spectrometer. The spectra were collected with a 40X objective and were dispersed through a grating of 2400 grooves·mm$^{-1}$. A thermoelectrically cooled charged coupled device (CCD) based detector was used for all these studies.

The morphological study was carried out by using a field emission scanning electron microscope (FESEM; AURIGA, Zeiss). In order to investigate the distribution of the constituent elements in the NWs, energy-filtered transmission electron microscopic (EFTEM) imaging was carried out using transmission electron microscope (FEI TITAN 80-300 (Cubed)). The NWs were dispersed in isopropyl alcohol and were transferred to Cu grids. EFTEM imaging was performed for the constituent elements (Al, Ga, and N) corresponding to the plasmon energy loss of each element of the NWs with an energy filter having a resolution of 0.6 eV.

**2.3 Gas Sensing Measurements**

The gas sensing experiments were carried out using a custom-made gas exposure facility system in a dynamic condition under continuous pumping using a oil free scroll pump (Edward XDS10). The sensing chamber consists of a double-walled stainless-steel chamber of volume 30 liters containing a proportional–integral–derivative (PID) controlled heater. The digital picture of the gas sensing setup is shown in the Fig. 1. The details of the sensing chamber and the measurement are described elsewhere.[11] In brief, as a first step, the sample with Au contact pad was mounted inside the sensing chamber on the PID controlled heater, which can go up to a maximum temperature of 500 ($\pm$1) °C. The pressure contacts to the sample were established (Fig. 1). Then the system was pre-evacuated to ~1x10$^{-2}$ mbar and was subsequently purged with UHP N$_2$ to avoid presence of other gases. The cycle was repeated few times. Two metallic probes in contact with the sample could measure the resistance of the sample with the help of a micro-Ohm meter (Agilent 34401). Finally, the controlled gas mixture (UHP N$_2$ and CH$_4$) was allowed into the sensing chamber through mass flow controllers. All the gases were passed through moisture trap and hence at fixed humidity which was on the drier side. The resistance of the sample was continuously monitored with respect to time with the help of LabVIEW programming. The response of the sample was recorded upon the gas exposure based on the mechanism of change in resistance, monitored using the micro-Ohm meter.

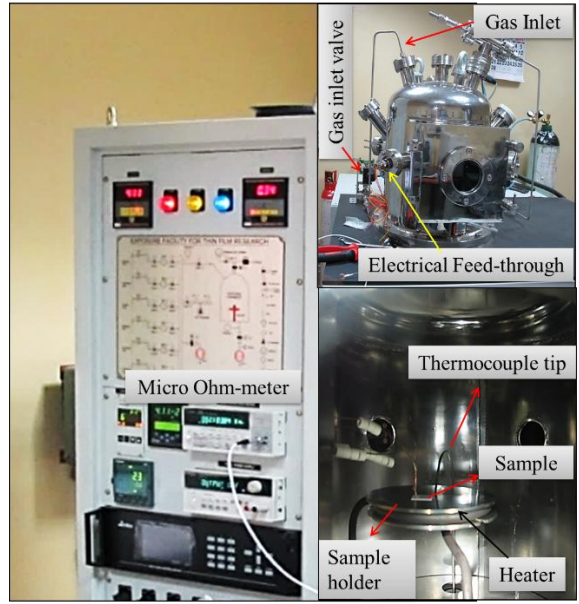

Fig. 1 The digital picture of the gas sensing setup with important components labeled

## 3. Results and discussion

### 3.1 Vibrational Analysis

Raman spectroscopic analysis using 514.5 nm excitation was carried out for the confirmation of the phase formation. The typical Raman spectra of as-grown GaN NWs and $Al_xGa_{1-x}N$ synthesized *via* IBM process are shown in the Fig. 2a. Corresponding FESEM images show (outset of Fig. 2a) NWs with average diameter ~100 nm. The IBM synthesized $Al_xGa_{1-x}N$ NWs preserve their size, and shape after the irradiation with high energy ions and subsequent annealing at 1000 °C. In case of as-grown GaN NWs, the Raman peaks at ~567, and ~725 cm$^{-1}$ correspond to the symmetry allowed $E_2$(high) and $A_1$(LO) modes, respectively, for wurtzite GaN.[26,27] Along with the symmetry allowed modes, another peak is observed at ~ 420 cm$^{-1}$, which may correspond to zone boundary phonon mode arising due to the finite crystal size of GaN NWs.[28] The peak centered at ~520 cm$^{-1}$ is originated from the crystalline Si substrate. Whereas, the spectrum for $Al_xGa_{1-x}N$ NWs shows (Fig. 2a) all the above discussed vibrational modes with comparatively low intensities. Since random alloy formation is expected for the $Al_xGa_{1-x}N$, the $E_2$(high) mode corresponding to AlN (AlN-$E_2$(high)) is also probable to be observed along with the $E_2$(high) mode of GaN. The absence of the AlN-$E_2$(high) mode may be owing to the nominal incorporation of Al percentage in GaN, which is not sufficient enough to be probed in the usual Raman spectroscopy. The $A_1$(LO) mode could not retrieve back the sufficient peak intensity in the Raman spectra. Resonance Raman spectroscopy (RRS), invoking the Fröhlich interaction, was performed to understand the behavior of the $A_1$(LO) mode with the help of 325 nm excitation laser source in the presence of strong electron-phonon coupling.[29,30]

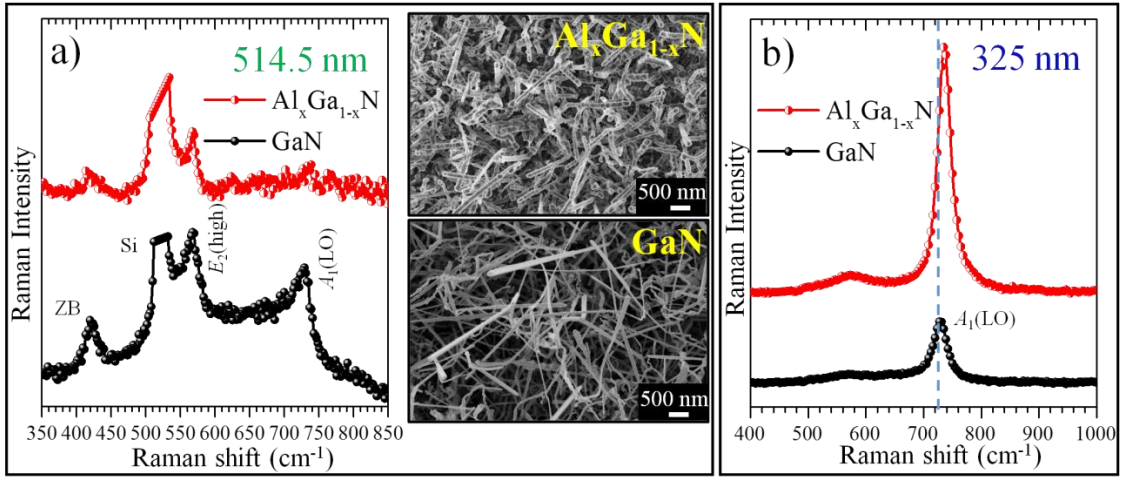

Fig. 2 (a) Typical Raman spectra of as-grown GaN and $Al_xGa_{1-x}N$ synthesized *via* IBM process. Outset shows typical FESEM micrograph of the NWs. (b) Corresponding Resonance Raman spectra probed with an excitation laser source of 325 nm. The vertical dashed line is a guide to eye for the blue shift of $A_1(LO)$ mode.

The typical RRS spectra of the GaN and $Al_xGa_{1-x}N$ are shown in the Fig. 2b. The $A_1(LO)$ mode of $Al_xGa_{1-x}N$ arises with a blue shift of ~10 $cm^{-1}$ confirming the one-mode behaviour of the random alloy.[27] According to the band bowing formalism of random alloy model, the percentage of Al incorporation is given by the following equation which is analogous to the Vegard's law.[25,27]

$$A_1(LO)_{AlGaN} = A_1(LO)_{GaN} + [A_1(LO)_{AlN} - A_1(LO)_{GaN}]x - bx(1-x) \quad \text{...........(1)}$$

where '*x*' is the Al atomic percentage in AlGaN and '*b*' is the bowing parameter. The bowing parameter '*b*' can be neglected for a lower percentage of Al. Thus, the atomic percentage of Al in the AlGaN random alloy is found to be ~7 %.

## 3.2 Structural and Elemental Analysis

The distribution of the constituent elements in the NWs is demonstrated with the help of EFTEM imaging. Fig. 3 depicts the EFTEM images of a single $Al_{0.07}Ga_{0.93}N$ nanowire imaged with the corresponding volume and/or surface plasmon energy edge of the constituent elements.

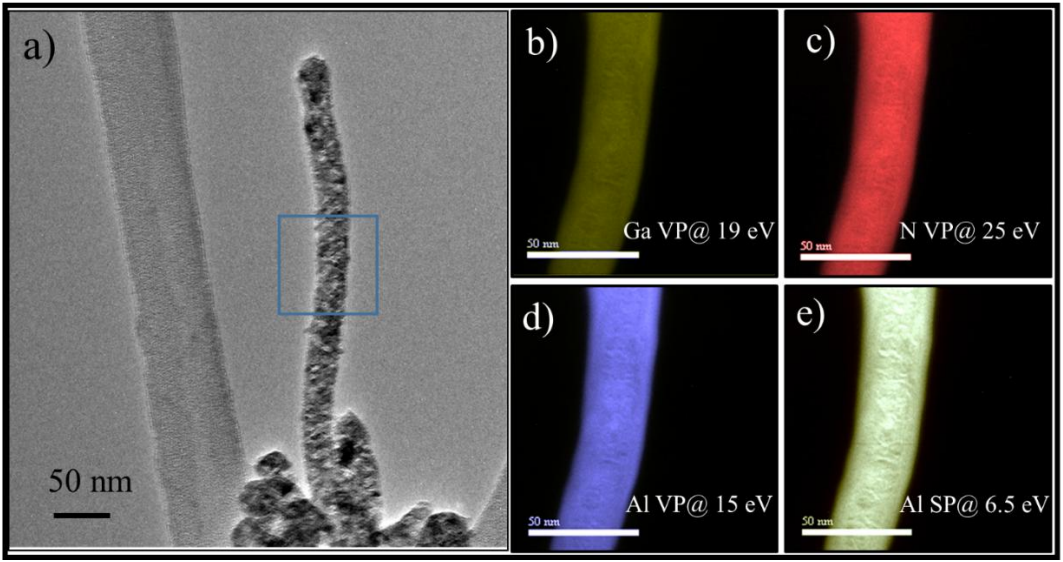

Fig. 3 (a) Transmission electron microscope image of a single Al$_{0.07}$Ga$_{0.93}$N nanowire. The corresponding energy-filtered transmission electron microscope image of the selected portion of the nanowire with (b) Ga (c) N and (d) Al volume plasmon edge, (e) Al surface plasmon edge.

The EFTEM image of a single nanowire (diameter ~30 nm) is taken from the marked rectangular area (Fig. 3a). The images recorded with the volume plasmon edges of Ga, N, and Al are shown in the Figs. 3b, 3c, and 3d, respectively. Similarly, the image with surface plasmon edge of Al is depicted in the Fig. 3e. The EFTEM images confirm the uniform distribution of the constituent elements throughout in the nanowire.

### 3.3 Gas sensing performance

The gas sensing of as-grown GaN and Al$_{0.07}$Ga$_{0.93}$N was carried out with exposing the sample to CH$_4$. The change in resistance of the samples was measured as a function of time for different concentrations of the analyte at different operating temperatures. The response curves of the GaN NWs with the exposure of CH$_4$ gas at temperatures of 150 °C (Fig. 4a) and 200 °C (Fig. 4b) are shown with a variation of concentration ranging from 100 to 500 ppm. Response (*S*) of the sensor is calculated as,

$S = \dfrac{R_0 - R_G}{R_0}$, where $R_0$ and $R_G$ are the resistance of the sample in the absence and presence of analyte gas, respectively.

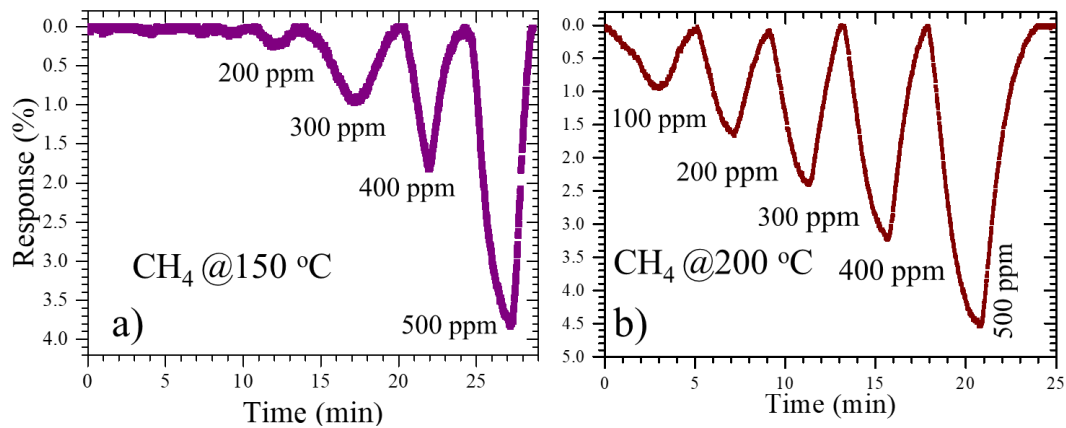

Fig. 4 CH$_4$ gas sensing response of GaN NWs at (a) 150 and (b) 200 °C with different gas concentration ranging from 100 to 500 ppm.

At 150 °C, the sample does not show a significant response to the 100 ppm of CH$_4$. However, it starts measurable response for 200 ppm with a response of ~0.25 % and increases up to ~3.8 % on the exposure of 500 ppm (Fig. 4a). With the increase in operating temperature to 200 °C, the GaN NWs even show the response for 100 ppm and the response of ~0.95 % at 100 ppm goes up to ~4.5 % by increasing the amount of exposure gas to 500 ppm (Fig. 4a). Increase in the response with temperature supports the chemisorption driven sensing process.[22]

The CH$_4$ gas sensing behaviour of Al$_{0.07}$Ga$_{0.93}$N, synthesized *via* IBM process, is shown in Figs. 5a-d. In contrast to GaN NWs, the sample starts responding at 50 °C even for the lowest amount of CH$_4$ exposure of 100 ppm (Fig. 5a). At 100 ppm the response is ~2.5 %, which increases up to ~6.0 % for the exposure of 500 ppm. The maximum response of ~20.0 % was recorded with an exposure of 500 ppm at the sample temperature of 150 °C (Fig. 5c) implying a ~4.5-fold increase in the response value over that of GaN NWs at 200 °C.

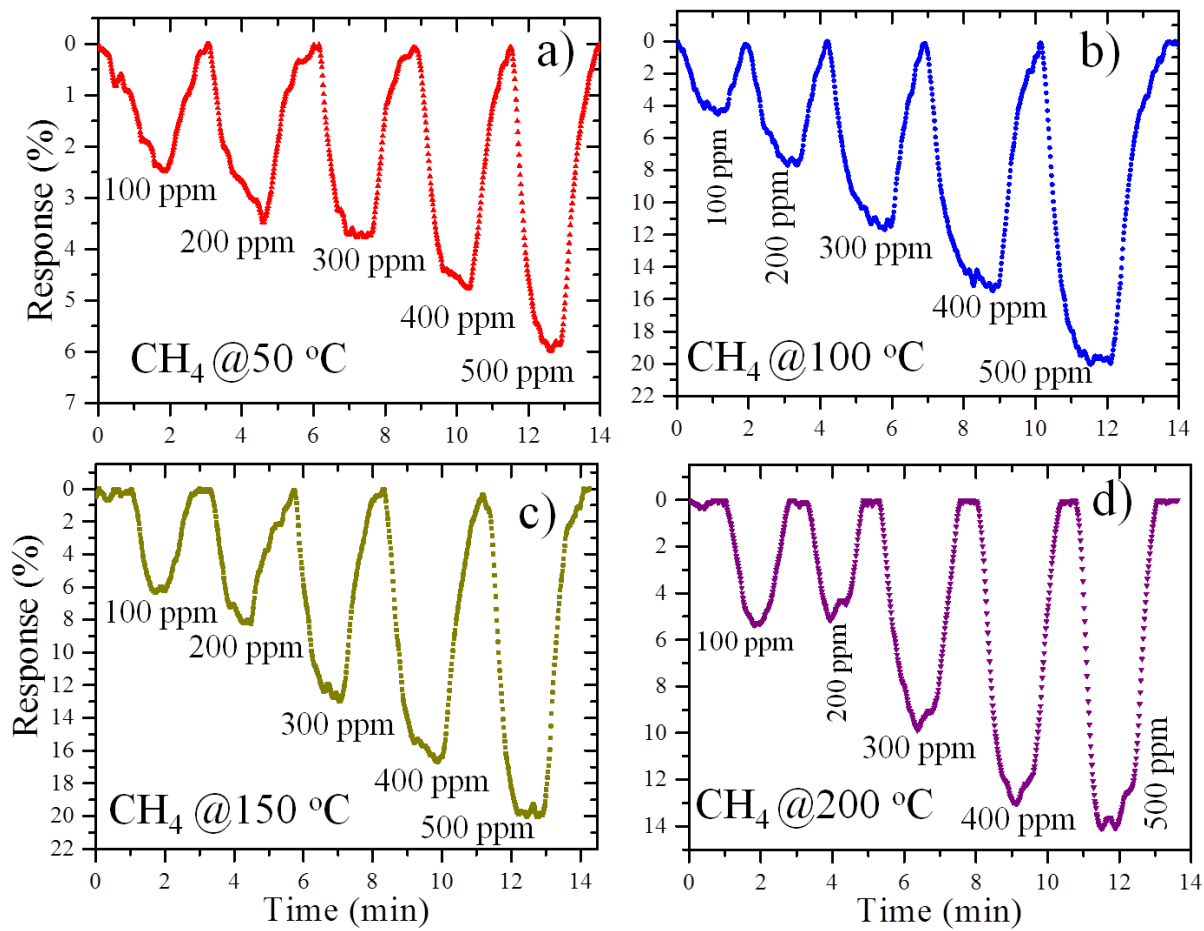

Fig. 5 CH$_4$ gas sensing response of Al$_{0.07}$Ga$_{0.93}$N NWs at (a) 50, (b) 100, (c) 150 and (d) 200 °C with different gas concentrations ranging from 100 to 500 ppm

The plots of variation in the CH$_4$ response at different operating temperatures and the analyte concentrations are presented (Fig. 6a) for Al$_{0.07}$Ga$_{0.93}$N NWs. The response increases with increase in the operating temperature up to 150 °C and decreases by further raising the temperature to 200 °C (Fig. 6a). The decrease in response is attributed to the reaction kinetics and the adsorption-desorption process of CH$_4$ for the chemisorption based sensor response. Once the desorption process takes over, the sensor response also drops, and the changes of response at 200 °C for different gas concentration also become random. These phenomena indicate conditions favoring CH$_4$ reaction in the temperature range of 100 to 150 °C, while desorption phenomena occur dominantly above 150 °C. However, the optimum response for GaN is observed at 200 °C. This difference in optimum response temperature can be attributed to the presence of the additive in the system.[31] Incorporation of additive or dopant in the semiconductor can generate active sites which leads to increase in surface energy.[31-33] Therefore, the kinetics of the reactions are affected leading to different adsorption and desorption rates at a particular temperature. In the present study, Al acts as an additive in GaN for the modification of the optimum sensing temperature. Notably, $Al_{0.07}Ga_{0.93}N$ NWs showed stable sensor response with high repeatability over a period of few months.

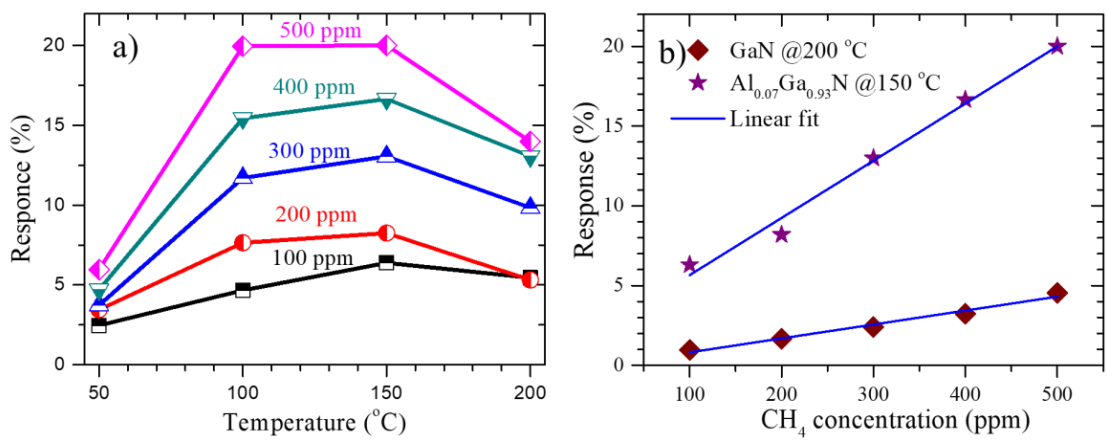

Fig. 6 (a) Variation of $CH_4$ gas response towards $Al_{0.07}Ga_{0.93}N$ with respect to temperature and analyte concentration. Solid lines are guide to eye for the variation. (b) Comparative sensing response of GaN and $Al_{0.07}Ga_{0.93}N$ as a function of $CH_4$ concentration at operating temperatures of 150 and 200 °C, respectively. Solid lines are linear fit for the recorded response.

Further, as a comparative study, the variation of sensing response as a function of analyte concentration is depicted (Fig. 6b) at the optimum operating temperatures of 200 °C for as-grown GaN and 150 °C for $Al_{0.07}Ga_{0.93}N$. Both the samples show nearly linear behaviour, signifying the sufficient number of available active sites even after the exposure of 500 ppm of $CH_4$. Interestingly, it can be easily inferred from Fig. 6b that in case of the $Al_{0.07}Ga_{0.93}N$, there is always a possibility of detecting further lower concentration (<100 ppm) of $CH_4$ with a significant response of ~4 %. In order to check the selectivity, we have also tested the response of $Al_{0.07}Ga_{0.93}N$ nanowires towards the oxidizing gas $O_2$, the second most abundant gas in the atmosphere after the inert $N_2$. However, there is no measurable change in the resistance of the $Al_{0.07}Ga_{0.93}N$ nanowires with the exposure $O_2$ (not shown in figure). Our earlier study with $NO_x$, $NH_3$, and $H_2$ also failed in the absence of metallic catalyst nanoparticle for GaN nanostructures.[6,15] The obtained $CH_4$ sensing response of the present study is compared with the existed literature for other materials and is presented in Table 1.[10,11,22,34-38]

Table 1. Comparison of CH$_4$ sensing response towards different materials.

| Material | Morphology | Concentration (ppm) | Sensing Temperature (°C) | CH$_4$ Response (%) | Reference |
|---|---|---|---|---|---|
| SnO$_2$ | Nanoparticle | 500 | 250 | 18 | 10 |
| VO$_2$ | Nanorod | 500 | 50 | 3 | 11 |
| SnO$_2$ | Nanoparticle | 400 | 225 | 6 | 22 |
| In doped CaZrO$_3$ /MgO composite | Porous structure | 1000 | 600 | 3 | 34 |
| Ca- impregnated SnO$_2$ | Micro structure | 1000 | 400 | 80 | 35 |
| ZnO | Nanoparticle | 100 | Room Temperature | 80 | 36 |
| Au decorated VO$_2$ | Nanosheet | 500 | Room Temperature | 70 | 37 |
| VO$_2$ | Nanorods | 500 | Room Temperature | 26.5 | 38 |
| **GaN** | **Nanowire** | **500** | **200** | **4.5** | **Present work** |
| **Al$_{0.07}$Ga$_{0.93}$N** | **Nanowire** | **500** | **50** | **20** | **Present work** |

The key parameters like response and recovery time are compared for the optimum operating temperature of 200 °C for as-grown GaN and 150 °C for Al$_{0.07}$Ga$_{0.93}$N with the analyte concentration of 100 ppm. The response and recovery time of Al$_{0.07}$Ga$_{0.93}$N is found to be ~33 and 42 s, respectively. Whereas, for GaN the response and recovery time are 126 and 132 s, respectively. Less response and recovery time in Al$_{0.07}$Ga$_{0.93}$N NWs as compared to the GaN NWs prove the former as a better material of choice along with the significant sensor response. The overall performance of Al$_{0.07}$Ga$_{0.93}$N nanowires in CH$_4$ sensing is also found to be remarkable in regard to its low temperature operation with significant response in the optimized condition as compared to the oxide based sensors (Table 1).

**3.4 Gas sensing mechanism**

The CH$_4$ gas sensing mechanism of GaN was described in an earlier report.[9] GaN NWs were found to respond for the reducing gas CH$_4$ because of the presence of native defect complexes (V$_{Ga}$-O$_N$ and 2O$_N$), which were supported by both experiment and density functional theory based *ab initio* calculations. The O antisite defect (O$_N$) creates the Ga vacancy (V$_{Ga}$) leading to the creation of most stable defect complexes in the form of 2O$_N$ and V$_{Ga}$-O$_N$.[9] It is well known that Al is more intended to react with O than Ga, driven by the respective free energy of formation.[39] Therefore, the probability of formation of O$_N$ defect in Al$_{0.07}$Ga$_{0.93}$N is enhanced as compared to that in GaN. Additionally, some absorbed oxygen on Al/GaN may diffuse inside the GaN along with Al in the process of ion beam irradiation. In this context, the trace amount of Al and O prefers to occupy the lattice sites of Ga and N, respectively as a dopant for the overall energy minimization of the crystal.[40] This process can lead to

`

the presence of a relatively large amount of charged $O_N$ defects in $Al_{0.07}Ga_{0.93}N$ NWs. In consequence, it augments the observation of higher response for the $Al_{0.07}Ga_{0.93}N$ NWs at low temperature than that for GaN NWs. It is further discussed in the end of this section. The formation of the oxide phase of Ga or Al in the system is, however, discarded due to the absence of any signature in the Raman spectra (Fig. 2). Therefore, a significant majority of the oxygen present in the system belongs to the defects in the form of point defects like $O_N$ and defect complexes of $2O_N$ and $V_{Ga}$-$O_N$. The $O_N$ defect in $Al_xGaN_{1-x}N$ is reported to possess negative charge,[40] which may take part in the sensing process. In order to further enrich our argument, the PL spectral study, a well-known characterization tool for defect analysis, was performed. The PL spectra were collected for both as-grown GaN and IBM $Al_{0.07}Ga_{0.93}N$ NWs at room temperature. We have also compared the PL data with those for the GaN sample irradiated and post-annealed without any Al coating to understand that defects present is unique to $Al_{0.07}Ga_{0.93}N$ phase and is not due to the irradiation effect alone (Fig. 7).

The luminescence from the as-grown GaN NWs shows broadband in the range of 3.0-3.6 eV (Fig. 7a). The peak at ~3.51 eV is assigned to free exciton (FE) corresponding to the emission due to the recombination of electron-hole (*e-h*) pairs.[41] The emission peak ~3.28 eV is also observed in the PL spectrum along with the FE emission. The peak at ~3.28 eV is attributed to the recombination of the donor-acceptor pair (DAP) for a transition of electrons from a shallow donor state ($V_N$, $O_N$) to a deep acceptor state of $V_{Ga}$.[24,41] The phonon replica of DAP transition is also observed in the PL spectra. In case of the GaN sample irradiated and post-annealed at 1000 °C, the PL spectrum (Fig. 7b) exactly follows the line shape and peak position as that of the as-grown GaN NWs. However, there is a decrease (~0.3 times) in the PL intensity for the irradiated sample, which can be attributed to the increase in the non-radiative centers by the process of irradiation. The PL spectrum of the $Al_{0.07}Ga_{0.93}N$ sample is de-convoluted to 3 peaks centered ~3.25, 3.11 and 2.94 eV (Fig. 7c). The luminescence peaks at ~3.25 and 3.11 eV are attributed to the DAP and the phonon replica of the DAP transition, respectively. These two peaks completely follow the line shape and nature of transition as observed in the former two samples of as-grown GaN and post-irradiation annealed GaN. The peak ~2.94 eV, arising with a significant intensity, can be attributed to the blue luminescence (BL) which is assigned to the transition between the defect states of deep donors ($V_N$) and shallow acceptors ($V_{Ga}$).[41] The schematic band diagram is shown (Fig. 8) for different defects states which are related to the observed luminescence (Fig. 7).

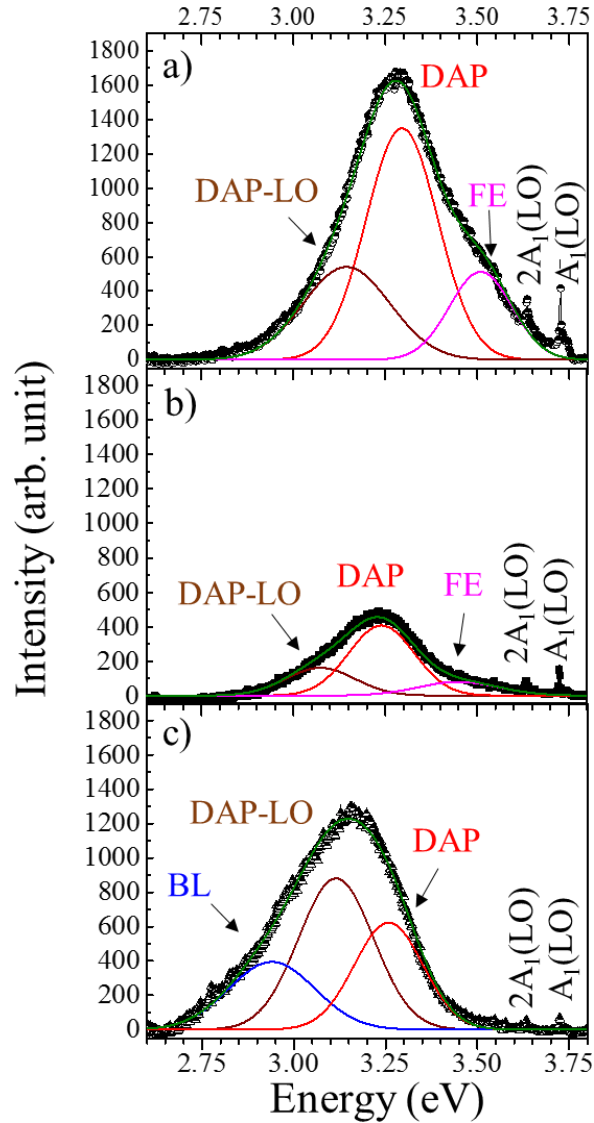

Fig. 7 Typical PL spectra of (a) as-grown GaN, (b) post-irradiated annealed GaN and (c) $Al_{0.07}Ga_{0.93}N$ NWs.

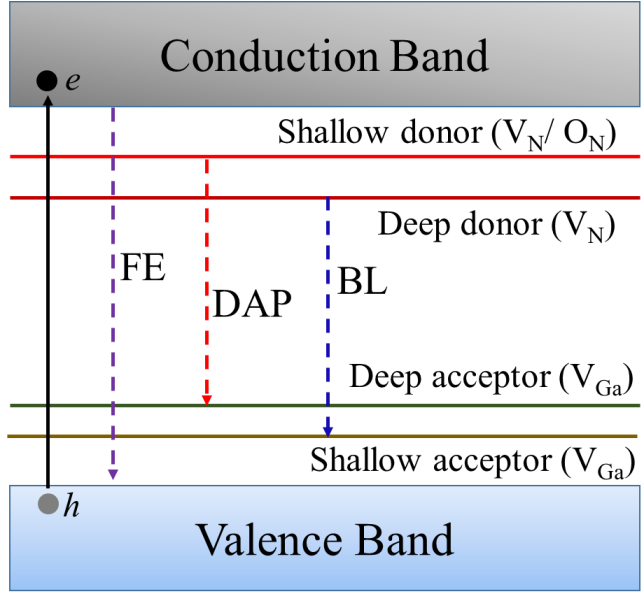

Fig. 8 Schematic band diagram with different defect states and corresponding luminescence.

The BL band is observed only in the $Al_{0.07}Ga_{0.93}N$ NWs. Considering the above fact, it can be concluded that the number of shallow acceptor-like $V_{Ga}$ defects are higher in $Al_{0.07}Ga_{0.93}N$ NWs than those in the as-grown GaN NWs. Since the post-irradiation annealed GaN sample does not show any BL band, these defects are not created during irradiation. Generally, the shallow acceptor states are produced in two ways by $V_{Ga}$ defect in GaN. Firstly, if $V_N$ is created in the system, there will be a formation of $V_{Ga}$ in order to maintain the net charge neutrality in the solid.[7] Similarly, $V_{Ga}$ can also be produced in the system, if O atoms sitting in the lattice sites of N ($O_N$), and renders a shallow acceptor state. Thus, the creation of $V_{Ga}$ with the incorporation of $O_N$ is due to the fact of maintaining the net charge neutrality in the system. The $V_{Ga}$, produced in the system because of the formation of the $V_N$, can be either a shallow or deep acceptor state.[42] Unlike the $V_N$, which can create deep or shallow donor states depending on the formation energy, $O_N$ always creates a shallow donor state just below the conduction band.[40,43,44] Hence, the $V_{Ga}$ formed in the system because of the anticipation of $O_N$ will always form a shallow acceptor state.[7] In the present study for $Al_{0.07}Ga_{0.93}N$ NWs, the BL band in the PL spectra is observed indicating the presence of shallow acceptor level which is created by the $V_{Ga}$. This observation emphasizes that the excess $V_{Ga}$ are created because of the abundance of $O_N$ over $V_N$. Hence, the number of $O_N$ defect is higher in $Al_{0.07}Ga_{0.93}N$ NWs as compared to the as-grown GaN NWs.

In the present study, there is a reaction involved in the sensing process of $CH_4$ with the adsorbed oxygen, present as a negatively charged oxygen defect ($O_N$) in the nanowires followed by possible reduction to $CO_2$ and transfer of electrons as charge carriers.

$$CH_4 \text{(gas)} + 4O^- \rightarrow CO_2 + 2H_2O + 4e^- \dots\dots\dots(2)$$

Since charge carriers (in the form of electrons) are generated by the process of above reaction, the resistance of the sample drops in the presence of $CH_4$. Once the flow of $CH_4$ is stopped into the sensing chamber, the sample retains back to its original resistance in the operating temperatures with a trace of oxygen available in the chamber. From the earlier discussion on PL spectroscopic analysis (Fig. 7), we have clear evidence that the number of negatively charged $O_N$ defect is higher in $Al_{0.07}Ga_{0.93}N$ NWs as compared to that in the as-grown GaN NWs. Therefore, $Al_{0.07}Ga_{0.93}N$ NWs help in the sensing of $CH_4$ with a higher response at a comparatively low temperature (50 °C) than that of GaN NWs.

## 4. CONCLUSIONS

Low operating temperature $CH_4$ gas sensing is realized in the AlGaN nanowires synthesized via ion beam mixing process. The vibrational analysis confirms the wurtzite phase of the random alloy in AlGaN with Al incorporation percentage of ~7 at.%. The $Al_{0.07}Ga_{0.93}N$ nanowires start responding to $CH_4$ gas at a comparatively low operating temperature of 50 °C, and the maximum response of ~20% at 500 ppm is obtained in the temperature range of 100 to 150 °C. An increase in response about five times is recorded for $Al_{0.07}Ga_{0.93}N$ nanowires than that of the GaN nanowires in the optimum condition of the respective systems. Enhanced sensing response at low operating temperature is obtained in the $Al_{0.07}Ga_{0.93}N$ nanowires in comparison to that of the GaN nanowires. The gas sensing efficiency in $Al_{0.07}Ga_{0.93}N$ nanowires is significantly influenced by the presence of negatively charged O antisite defect. Thus, the present study provides a pathway for utilization of III-V nitrides for gas sensing.

## Acknowledgements

We thank P. Magudapathy of MPD, IGCAR for his help in carrying out the ion irradiation process. We acknowledge R. Pandian of SND, IGCAR, for his help in the FESEM study. We also extend our thanks to A. Patsha, Kishore K. Madapu, and A. K Sivadasan of SND, IGCAR, and S. Bera of Water and Steam Chemistry Division, BARC Facility at Kalpakkam for their valuable suggestions and useful discussions.

`

`